\documentclass[fleqn,english,aps,prd,preprintnumbers,showpacs,showkeys,nofootinbib,superscriptaddress,floatfix,tightenlines]{revtex4-2}
\usepackage[T1]{fontenc}
\usepackage[utf8]{inputenc}
\usepackage{amsmath}
\usepackage{amsthm}
\usepackage{amssymb}
\usepackage{graphicx}
\usepackage{amsfonts}
\usepackage{amscd}
\usepackage{amsxtra}
\usepackage{epstopdf}
\usepackage{color}
\usepackage{dcolumn}
\usepackage{bm}
\usepackage{multirow}
\usepackage{slashed}

\makeatletter

\usepackage[normalem]{ulem}
\usepackage[dvipsnames]{xcolor}

\renewcommand{\sout}{\bgroup \color{red} \ULdepth=-.5ex \ULset}

\makeatother

\normalsize

\usepackage{babel}
\begin{document}
\preprint{INHA-NTG-02/2024}
\title{Properties of neutron stars and strangeness-mixed stars from a
  pion mean-field approach} 
\author{Nam-Yong Ghim}
\email{Namyong.gim@inha.edu}
\affiliation{Department of Physics, Inha University, Incheon
22212, Republic of 
   Korea} 

  \author{Hyun-Chul Kim}
  \email{hchkim@inha.ac.kr}
\affiliation{Department of Physics, Inha University, Incheon
22212, Republic of Korea} 
\affiliation{School of Physics, Korea Institute for Advanced
Study (KIAS), Seoul 02455,  Republic of Korea}

\author{Ulugbek Yakhshiev}
\email{yakhshiev@inha.ac.kr}
\affiliation{Department of Physics, Inha University, Incheon
22212, Republic of Korea}
\affiliation{Theoretical Physics Department, National University of
  Uzbekistan, Tashkent 100174,  Uzbekistan} 

  \author{Ghil-Seok Yang}
  \email{ghsyang@hoseo.edu}
\affiliation{Department of General Education for Human
Creativity, Hoseo University, Asan 31499, Republic of Korea}  

\begin{abstract}
We investigate the properties of the static neutron stars and
strangeness-mixed stars, based on the equations of state derived from
a pion mean-field approach. Using the empirical data on the
pion-nucleus scattering and bulk properties of nuclear matter, we have
already fixed all the parameters in a previous work, where the
nucleons and hyperons were shown to be modified in various nuclear
medium. In the current work, we first examine the energy and pressure
inside a neutron star. We show that the central densities in various
neutron stars vary within the range of $(3-4)\rho_0$, where $\rho_0$
is the normal nuclear matter density. The mass-radius relations are
obtained and discussed. As the slope parameter for neutron matter
increases, the radii of the neutron stars increase with their masses
fixed. We also study the strangeness-mixed stars or the hyperon stars
using the same sets of the parameters. As the strangeness content of
strange matter increases, the binding energy per nucleon is saturated
and the corresponding equation of state becomes
softened. Consequently, the central densities of the strangeness-mixed
stars increase. Assuming that recently observed neutron stars are the
strangeness-mixed ones, we find that the central densities
increase.    

\end{abstract}
\maketitle 

\section{Introduction}
Neutron stars are known to form strongly interacting cold-dense
neutron matter, so that they provide a test ground for the equation of
state (EoS) derived from many theories and
models~\cite{Lattimer:2012nd, Oertel:2016bki}. The data on the masses
and radii of the neutron stars have been compiled from astronomical
observations such as low mass X-ray binaries~\cite{Paradijs:1979,
  Lewin:1993,   Rutledge:1998dt, Rybicki:2005id, Webb:2007tc,
  Steiner:2015aea, Alvarez-Castillo:2016oln, Ozel:2016oaf} for
decades. Demorest et al.~\cite{Demorest:2010bx} observed the binary
millisecond pulsar J1614-2230 from which its mass was estimated to be
$(1.97\pm 0.04)M_{\odot}$.  Antoniadis et
al.~\cite{Antoniadis:2013pzd} made a 
radio-timing observations of the PSR J0348+0432 and evaluated the mass
of the neutron star as $(2.01\pm 0.04)M_{\odot}$. Fonseca et
al.~\cite{Fonseca:2016tux} analyzed 24 binary radio pulsars for nine
years and found the mass range of the neutron stars between
$1.18\,M_{\odot}$ and $1.93\,M_{\odot}$. In a recent
paper~\cite{Fonseca:2021wxt}, the pulsar mass was estimated to 
be $2.08\,M_{\odot}$ from the observations of PSR J0740+6620. 
Very recently, the observation of gravitational waves have
further sharpened the data on the neutron 
stars~\cite{LIGOScientific:2016emj, LIGOScientific:2016sjg,
  LIGOScientific:2018hze}. The proposed Einstein Telescope in
Europe ~\cite{Punturo:2010zz, Maggiore:2019uih, Iacovelli:2023nbv} and
the Cosmic Explorer in the US~\cite{Reitze:2019iox, Evans:2021gyd},
which are ten times sensitive future third-generation
gravitational-wave observatories, are expected to provide more precise
information on the nature of neutron stars in the future.
The Neutron Star Interior Composition Explorer (NICER) experiment
observed PSR J0030+0451, from which the radius and mass of the neutron
star were inferred to be $R=13.02_{-1.06}^{+1.24}$ km and 
$M=1.44_{-0.14}^{+0.15}\,M_{\odot}$~\cite{Miller:2019cac}. On the 
other hand, Riley et al.~\cite{Riley:2019yda} estimated them as
$12.39_{-0.98}^{+1,30}$ km and $2.072_{-0.066}^{+0.067}\, M_{\odot}$. 
We anticipate that the NICER will bring more essential
data~\cite{Miller:2021qha, Riley:2021pdl}, which can further constrain
the EoS for neutron star matter (see also recent reviews and
references therein~\cite{Lattimer:2015nhk, Baym:2017whm}).
The mass of black widow~\cite{Romani:2022jhd}, which is the fastest
and heaviest known galactic neutron star, was estimated to be
$M=(2.35\pm 0.17) M_{\odot}$. Its observation can also constrain the
EoS. 

In the present work, we aim at investigating the masses and radii of
the neutron stars, based on the pion mean-field approach or the chiral 
quark-soliton model ($\chi$QSM)~\cite{Diakonov:1987ty, 
  Christov:1995vm, Diakonov:1997sj}. The idea was motivated by
Witten's seminal paper~\cite{Witten:1979kh}: In the large $N_c$ (the
number of colors) limit the presence of $N_c$ valence quarks creates
the pion mean fields self-consistently. Since the mesonic quantum
fluctuations are suppressed in the large $N_c$ limit, a baryon emerges
as a state of $N_c$ valence quarks bound by the pion mean fields. This
approach was successful in describing many properties of the
low-lying baryons such as mass splittings~\cite{Blotz:1992pw,
  Yang:2010id, Yang:2010fm}, static properties~\cite{Kim:1995bq,
  Kim:1996vk, Kim:1997ip, Kim:1999uf, Ledwig:2010tu, Yang:2015era,
  Yang:2018idi}, form factors~\cite{Kim:1995mr, Silva:2001st,
  Silva:2005fa, Kim:2019gka, Won:2022cyy}, parton distribution
functions~\cite{Diakonov:1996sr, Diakonov:1997vc}, and so on.
A great merit of the pion mean-field approach lies in the fact that
both the light and singly heavy baryons can be explained on an equal
footing. Replacing a light valence quark with a heavy quark, we find
that the singly heavy baryon arises as the bound state of the $N_c-1$
valence quarks by the pion mean fields. A heavy quark can be regarded
as a static color source in the limit of the infinitely large mass of
the heavy quark~\cite{Yang:2016qdz, Kim:2017khv, Yang:2018uoj,
  Kim:2018xlc, Kim:2018nqf, Yang:2019tst, Kim:2019wbg, Kim:2020nug,
  Suh:2022ean, Suh:2022atr} (see also a recent
review~\cite{Kim:2018cxv}).   

The pion mean-field approach can also be extended to the investigation
of nuclear matter and medium modification of baryons in it. While one
can introduce the quark chemical potential and examine 
quark matter~\cite{Carter:1998ji,Berger:1996hc}, it is difficult to
associate quark matter with nuclear matter directly. Thus, we employ a
variational approach that was successfully used in the medium modified Skyrme
models~\cite{Rakhimov:1996vq, Meissner:2006id, Meissner:2007ya,
  Yakhshiev:2010kf, Kim:2012ts, Yakhshiev:2015noa}. A virtue of the
variational approach is that the dynamical parameters can be fixed by
considering low-energy pion-nucleus scattering and reproducing
saturation and bulk properties of homogeneous nuclear matter. In fact,
the general structure of the flavor SU(3) collective Hamiltonian arises
from the embedding of the SU(2) soliton into
SU(3)~\cite{Witten:1983tw, Witten:1983tx}. Thus, the collective
Hamiltonian can be constructed based on $\mathrm{SU(2)}_T\otimes
\mathrm{U(1)}_Y$ symmetry~\cite{Yang:2010fm,Yang:2010id}. Thus, while 
we derive the collective Hamiltonian, we fix the dynamical parameters
by reproducing the properties of nuclear matter. Recently, we studied
the medium modification of the low-lying SU(3) baryons in four
different cases, i.e. symmetric matter, asymmetric matter, neutron
matter, and strange baryonic matter consistently, based on a pion
mean-field approach~\cite{Ghim:2021odo}. We were able to explain the
properties of nuclear matter such as the binding energy per nucleon,
symmetric energy, pressure, and medium
modifications of the low-lying SU(3) baryons and  
singly heavy baryons~\cite{Ghim:2022zob}. Thus, we want to examine the
masses and radii of neutron stars, using the EoS developed in
Refs.~\cite{Ghim:2021odo, Ghim:2022zob}.
This approach has two significant virtues: 
First, since we have already fixed all necessary dynamical parameters,
we do not need to perform any additional fitting procedure to evaluate
the masses and radii of neutron stars, which will indeed play a role
of the touchstone for the EoS developed in the pion mean-field
approach. Second, the medium modifications of the nucleon were
consistently considered.

Since we assume that neutron stars have already been cooled and 
stationary, we use the Tolman-Oppenheimer-Volkoff (TOV) equation
for stationary neutron stars. In addition, we need to
consider the effects due to the presence of
leptons~\cite{Bahcall:1965zz, Boguta:1981mw, Lattimer:1985zf,
  Lattimer:1991nc, Steiner:2004fi, Potekhin:2015qsa, Sharma:2015bna,
  Burgio:2021vgk}, which brings about the contribution from the
protons, in particular, at the crust of neutron
stars. This will come into play to restrict the
EoS~\cite{Fortin:2016hny}.  

The current work is organized as follows: 
In Sec. II, we recapitulate the pion mean-field approach, focusing on 
the collective Hamiltonian and a medium modification of
the model using the variational approach. 
In Sec. III, we examine the mass, energy, and pressure densities,
which will be used in the TOV equation. We will also show how we can
include leptonic contributions. 
In Sec. IV, we discuss the results for the masses and radii of the
neutron stars, and then we extend the formalism to investigate the
hybrid stars, employing the strange matter. 
In Sec. V, we draw conclusions and summarize the current work. 
\section{Pion mean-field approach}
\label{sec:2}
The $\chi$QSM is based on the low-energy QCD partition function
in Euclidean space 
\begin{align}
\mathcal{Z}_{\mathrm{eff}} &= \int D\psi D\psi^\dagger D\pi^a
                             \exp\left[ \int 
\psi^\dagger( i\rlap{/}{\partial} + i MU^{\gamma_5} + i \hat{m})\psi 
\right]= \int D\pi^a \exp\left( -S_{\mathrm{eff}}[\pi^a]\right),   
\label{eq:1}
\end{align}
where $\psi$ and $\psi^\dagger$ denote the quark fields
and $\pi^a$ stand for the pseudo-Goldstone
fields. $S_{\mathrm{eff}}[\pi^a]$ is the effective chiral action defined by 
\begin{align}
S_{\mathrm{eff}}[\pi^a] := -N_c \mathrm{Tr}\log\left( i\rlap{/}{\partial}
  + i MU^{\gamma_5} + i \hat{m} \right),  
\end{align}
where $M$ represents the dynamical quark mass. It is originally the
momentum-dependent one, which was derived from the fermionic zero
mode of individual instantons~\cite{Diakonov:1985eg,
  Diakonov:2002fq}. For simplicity, we turn off the momentum
dependence of $M$ and introduce the regularization to tame the quark
fields. $U^{\gamma_5}$ denotes the chiral field expressed as 
\begin{align}
U^{\gamma_5}(x) = \exp{\left(i\pi^a(x) \lambda^a \gamma_5\right)} =
  U(x)\frac{1+\gamma_5}{2} + U^\dagger (x) \frac{1-\gamma_5}{2}  
\end{align}
with $U(x) = e^{i\pi^a(x) \lambda^a}$. $\hat{m}$ designates the
current quark mass matrix written as 
\begin{align}
\hat{m} =
  \begin{pmatrix}
    m_{\mathrm{u}} & 0 & 0 \\ 0 & m_{\mathrm{d}} & 0 \\ 0 & 0 &
    m_{\mathrm{s}} 
  \end{pmatrix}
= m_0 \mathbf{1} + m_3 \lambda^3 + m_8 \lambda^8 
\end{align}
with 
\begin{align}
m_0 = \frac{m_{\mathrm{u}} + m_{\mathrm{d}} + m_{\mathrm{s}}}{3},\;\;
m_3 = \frac{m_{\mathrm{u}} - m_{\mathrm{d}}}{2},\;\;
m_8 = \frac{m_{\mathrm{u}} + m_{\mathrm{d}} -2 m_{\mathrm{s}}}{2\sqrt{3}}  .
\end{align}
Here, $\lambda^a$ represent the Gell-Mann matrices for the flavor
SU(3) group and $m_{\mathrm{u}}$, $m_{\mathrm{d}}$, and
$m_{\mathrm{s}}$ denote the current up, down, and strange quark
masses, respectively. The effective chiral action contains all orders
of the effective chiral Lagrangians with the low-energy 
constants, which can be derived from the gradient
expansion~\cite{Diakonov:1987ty, Choi:2003cz}.  

The classical nucleon mass or the chiral soliton mass can be obtained
by solving the nucleon correlation function. By solving the
correlation function, we mean that the $N_c$ valence quarks are
positioned in the lowest level bound by the pion mean fields produced
by the presence of the $N_c$ valence quarks self-consistently. Since
the mesonic quantum fluctuations are known to be suppressed in the
large $N_c$ limit~\cite{Witten:1979kh}, we can integrate over the
pseudo-Goldstone boson fields in Eq.~\eqref{eq:1} around the saddle
point $U_c(\bm{x})$. The equation of motion for the quarks with the
classical pion fields can be minimized by the Hartree approximation,
we obtain the pion mean fields with hedgehog symmetry 
\begin{align}
U_c(\bm{x}) = \exp\left(i\bm{\tau}\cdot \bm{n} P(r)\right),  
\end{align}
where $P(r)$ is called the profile function. 
In the current work, we will not follow the self-consistent procedure
to determine the profile functions and classical soliton
mass. Instead, we will maximally use the $\mathrm{SU(2)}_T\otimes
\mathrm{U(1)}_Y$ symmetry that arises from the embedding of the SU(2)
chiral soliton into SU(3) 
\begin{align}
U(\bm{x}) = 
  \begin{pmatrix}
U_c(\bm{x}) & 0 \\
0 & 1    
  \end{pmatrix}.
\end{align}
The dynamical parameters will be determined by using the experimental
and empirical data. This method is often called ``\emph{a
model-independent}'' approach for the chiral soliton, which has a
virtue when it is extended to the investigation of bulk properties of
nuclear matter and baryon properties in it.
Since the classical soliton does not carry quantum numbers of 
baryons, we need to quantize it. As mentioned already, the quantum
fluctuations of the pion fields are suppressed by the large $N_c$
approximation. On the other hand, the fluctuation of the pion field to
the zero-mode directions, which are always related to translational
and rotational symmetries, must be completely treated. Rotational zero
modes determine the quantum numbers of baryons such as the spin and
isospin. For details, we refer to Refs.~\cite{Christov:1995vm,
  Diakonov:1997sj}.  

Having performed the rotational zero-mode quantization, we derive the
collective Hamiltonian~\cite{Blotz:1992pw, Yang:2010fm, Yang:2010id}
\begin{align}
H  = M_{{\rm cl}}
\;+\;H_{\mathrm{rot}}
\;+\;H_{\mathrm{sb}} + \; H_{\mathrm{em}},
\label{eq:7}
\end{align}
where $M_{\mathrm{cl}}$ denotes the mass of the classical
nucleon. $H_{\mathrm{rot}}$ stands for the rotational $1/N_c$
corrections arising from the zero-mode quantization
\begin{align}
H_{\mathrm{rot}} 
& =  
\frac{1}{2I_{1}}\sum_{i=1}^{3} \hat{J}_{i}^{2}
\;+\;\frac{1}{2I_{2}}\sum_{p=4}^{7} \hat{J}_{p}^{2},
\label{eq:8}
\end{align}
where the $I_{1,2}$ are the moments of inertia and  
$\hat{J}_a$ ($a=1,\cdots 8$) are the generators of the
$\mathrm{SU(3)}$ group, which are directly related to the right
angular velocities $\Omega_a$. The eighth component of $J_a$ is
constrained to be
\begin{align}
  \label{eq:9}
J_8 = -\frac{N_c}{2\sqrt{3}} B = -\frac{\sqrt{3}}{2},  
\end{align}
where $B$ is the baryon number. $J_8$ is related to the right
hypercharge  $Y_R=-Y'=(2/\sqrt{3})J_8$. While in the Skyrme model
$J_8$ is constrained by the Wess-Zumino term, the presence of the
$N_c$ valence quarks constrained it within the current approach.
The constraint \eqref{eq:9} has significant physical implications: It
selects the lowest allowed representations. If the baryon multiplet
with $Y=1$ consists of  $2J+1$ multiplet, the spin of the multiplet is
set to be $J$. Thus, the baryon octet and decuplet are selected. 
In the representation $\mathcal{R}=(p,q)$, the eigenvalues of
the Hamiltonian $H_{\mathrm{rot}}$ in Eq.~\eqref{eq:8} are given as  
\begin{align}
  E_{(p,\,q),\,J} & =  
  \frac{1}{2}
\left(\frac{1}{I_{1}}\,-\,\frac{1}{I_{2}}\right)J(J+1)-\frac{3}{8I_{2}}
+\frac{1}{6I_{2}}\Big(p^{2}\,+\,q^{2}\,+\,3(p\,+\,q)+\,p\,q\Big).
  \label{eq:10}
  \end{align}

The third term in Eq.~\eqref{eq:7} stands for both the isospin and
SU(3) flavor symmetry breaking part
\begin{align}
H_{\mathrm{sb}} & =  
\left(m_{\mathrm{d}}-m_{\mathrm{u}}\right)
\left(\frac{\sqrt{3}}{2}\,\alpha\,D_{38}^{(8)}(\mathcal{A})
\;+\;\beta\,\hat{T_{3}}\;+\;\frac{1}{2}
\,\gamma\sum_{i=1}^{3}D_{3i}^{(8)}(\mathcal{A})
\,\hat{J}_{i}\right)
\cr
 &   
+\;\left(m_{\mathrm{s}}-\bar{m}\right)
\left(\alpha\,D_{88}^{(8)}(\mathcal{A})
\;+\;\beta\,\hat{Y} +\;\frac{1}{\sqrt{3}}
\,\gamma\sum_{i=1}^{3}D_{8i}^{(8)}(\mathcal{A})
\,\hat{J}_{i}\right),
\label{eq:Hsb}
\end{align}
where $\alpha$, $\beta$, $\gamma$ are expressed in terms of the
moments of inertia
\begin{align}
\alpha
&=-\left(\frac{2}{3}
\frac{\Sigma_{\pi N}}{m_{\mathrm{u}}+m_{\mathrm{d}}}
-\frac{K_{2}}{I_{2}}\right),
\;\;\;\;
\beta=-\frac{K_{2}}{I_{2}},\;\;\;\;
\gamma=2\left(\frac{K_{1}}{I_{1}}-\frac{K_{2}}{I_{2}}\right).
\label{eq:abg}
\end{align}
Here $K_{1,2}$ represents the anomalous
moments of inertia of the rotating soliton.
The $\overline{m}$ is the average value of the 
up and down current quark masses, i.e. $\overline{m}=(m_{\mathrm{u}}+
m_{\mathrm{d}})/2$. The $D_{ab}^{(\mathcal{R})}(\mathcal{A})$ are the 
SU(3) Wigner $D$ functions in a given representation
$\mathcal{R}$. The explicit form of the electromagnetic (EM)
self-energy term $H_{\mathrm{em}}$ is given in
Ref.~\cite{Yang:2010id}. However, the effects of the EM self-energies  
are negligibly small for medium modification~\cite{Meissner:2006id,
  Meissner:2007ya}, we ignore them in the current work. 

The dynamical parameters $\alpha$, $\beta$, and $\gamma$ can be
evaluated within the self-consistent SU(3) $\chi$QSM. However, we will
adopt the model-independent approach, as previously mentioned. So, we
will fix all dynamical parameters by using the experimental data
rather than by computing within the model~\cite{Adkins:1984cf,
  Yang:2010fm}. Furthermore, we employ the variational method to
modify the collective Hamiltonian~\eqref{eq:7} in nuclear medium. We
briefly recapitulate the general formalism for the medium modification
of $H$. For details, we refer to
Refs.~\cite{Ghim:2021odo,Ghim:2022zob}.  

Before we proceed to discuss the density dependence of the binding
energy per nucleon, we will briefly mention the corrections from
the translational zero modes that make the nucleon carry the linear 
momentum. The zero-mode translation can be included by translating the
chiral soliton as 
  \begin{align}
    U_x(\bm{x}-\bm{X}(\tau)) = T_{\bm{X}(\tau)} U_c(\bm{x})
    T_{\bm{X}(\tau)}^\dagger,
  \end{align}
where $\bm{X}(\tau)$ is the translational zero modes conjugate to the 
linear momentum of the soliton. $T_{\bm{X}(\tau)}$ is a unitary SU(2)
matrix that governs the corresponding translation. Once we carry out
the quantization by expanding the effective chiral action with respect
to the velocity $\dot{\bm{X}(\tau)}$ to the second order, and
compute the functional integration over translational zero modes, we
have kinetic corrections to the collective Hamiltonian given in
Eq.~\eqref{eq:7} 
\begin{align}
H' = H + \frac{\hat{\bm{P}}^2}{2M_{\mathrm{cl}}},
\end{align}
where the kinetic term is of the order $\mathcal{O}(N_c^0)$ in the
large $N_c$ limit. $\hat{\bm{P}}$ is the linear momentum operator
conjugate to $\hat{\bm{X}}$.  
However, there is one important caveat. The classical mass
$M_{\mathrm{cl}}$ already contains spurious contributions from the
center-of-mass motion, which must be subtracted. This process is
nontrivial as was discussed in depth within the Skyrme
models~\cite{Moussallam:1993jd, Holzwarth:1994uj}, with $1/N_c$
pion-loop corrections included. Technically, this has resemblance with 
the random phase approximation in many-body theory. We may expect that 
two main subtraction terms expressed as 
\begin{align}
\Delta E = - \frac{\langle P^2\rangle}{2M_{\mathrm{cl}}} -
  \frac{\langle J^2\rangle}{2I},   
\end{align}
where $\langle p^2\rangle$ and $\langle J^2\rangle$ denote the average
values of $P^2$ and $J^2$. The derivation of these values is beyond the
current theoretical framework. Thus, we assume that the classical mass
of the nucleon effectively carry the corrections from the kinetic energy.

In Refs. ~\cite{Ghim:2021odo,Ghim:2022zob}, three different types of
nuclear environments were considered, i.e. the isospin symmetric,
isospin asymmetric, and strangeness-mixed nuclear matters.
Thus, we introduce three corresponding parameters as follows: 
\begin{align} 
\lambda=\frac{n}{n_{0}}, \quad  \delta
       =\frac{N-Z}{A},\quad  \delta_{s} 
       =\frac{N_{s}}{A} ,
  \label{eq:delta,lambda}
\end{align}
where $\lambda$ is the density of nuclear matter normalized by the
normal nuclear matter density $\rho_0\simeq 2.7\times
10^{14}\,\mathrm{g/cm^3}$ ($n_0=0.16\,\mbox{nucleons per
  $\mathrm{fm}^3$}$). The isospin asymmetry in nuclear matter is
controlled by $\delta$, which is related to the number of neutrons
$N$, of protons $Z$, and of baryons $A$.  The last variational
parameter $\delta_s$ is responsible for the strangeness mixing ratio
that is proportional to the number of baryons $N_{s}$ with strangeness 
$s=|S|$.  Using these three parameters, we can define the binding
energy per baryon as
\begin{align}
\varepsilon(\lambda,\delta,\,\delta_{1},
  \delta_{2},\delta_{3})    
&=
\Delta M_{N}\left(\lambda,\delta,\delta_{1},
    \delta_{2},\delta_{3}\right)
    \left(1-\sum_{s=1}^{3}\delta_{s}\right) +\frac{1}{2}\delta\,\Delta M_{np}
    \left(\lambda,\delta,\delta_{1},
    \delta_{2},\delta_{3}\right) 
    \cr 
&+\sum_{s=1}^{3}\delta_{s}\Delta M_{s}\left(\lambda,
    \delta,\delta_{1},\delta_{2},\delta_{3}\right).
\label{eq:BE}
\end{align}
See Ref.~\cite{Ghim:2021odo} for details. As a result,
we have the following density-dependent classical nucleon mass
$M_{\mathrm{cl}}^{\ast} $, the moments of inertia
$I_{\mathrm{1,2}}^{\ast}$, the in-medium modified
isospin symmetry breaking\footnote{Here, the strong part of the in-medium
  modified effects on isospin symmetry breaking are presented.} 
$E_{\mathrm{iso}}^{\ast}$ and the $\mathrm{SU(3)}$ symmetry breaking $E_{\mathrm{str}}^{\ast}$
terms:
\begin{align}
M_{\mathrm{cl}}^{\ast}  &= 
M_{\mathrm{cl}}f_{\mathrm{cl}}(\lambda,\delta,\delta_1,\delta_2,\delta_3),
\\
I_{\mathrm{1}}^{\ast}  &=  
I_{\mathrm{1}}f_1(\lambda,\delta,\delta_1,\delta_2,\delta_3),
\\
I_{\mathrm{2}}^{\ast}&=
I_{\mathrm{2}}f_2(\lambda,\delta,\delta_1,\delta_2,\delta_3),
\\
E_{\mathrm{iso}}^{\ast} &= 
\left(m_{\mathrm{d}}-m_{\mathrm{u}}\right)
\frac{K_{1,2}}{I_{1,2}^*}f_0(\lambda,\delta,\delta_1,\delta_2,\delta_3),
\label{eq:isosm}
\\
E_{\mathrm{str}}^{\ast} &=  
\left(m_{\mathrm{s}}-\bar{m}\right) 
\frac{K_{1,2}}{I_{1,2}^*}f_{\mathrm{s}}(\lambda,\delta,\delta_1,\delta_2,\delta_3), 
\label{eq:strm}
\end{align}
where $f_{\mathrm{cl}}$, $f_{\mathrm{0,1,2}}$, and $f_{s}$ are
given as the functions of the baryon environment densities and other
medium parameters. These functions are explicitly written as
\begin{align}
f_{\mathrm{cl}}(\lambda)
& =\left(1 + C_{\mathrm{cl}}\lambda\right),
\\
  f_{1,2}(\lambda)
& =\left(1 + C_{1,2}\lambda\right),
\label{eq:den_ftcl12}
\\
f_0(\lambda,\delta)
& = 1 + \frac{C_{\mathrm{num}}\lambda\,\delta}{1 + C_{\mathrm{den}}\lambda}\,,
  \label{eq:f0}
\\
  f_{s}(\lambda,\delta_{s})
& = 1 + g_{s}(\lambda)\delta_{s}
\\
  g_s(\lambda) 
&=  sg(\lambda),
\\
g(\lambda)
& = \left(6\frac{K_2}{I_2} + \frac{K_1}{I_1}\right)^{-1}
\cr
& \hspace{-1cm} \times 
  \frac{5(M_{cl}-M_{\mathrm{cl}}^{\ast} +
       E_{(1,1)1/2}- E_{(1,1)1/2}^*)}{3(m_s-\hat m)},
\label{eq:gs}
\end{align}
where the $C_{\mathrm{cl}}$, $C_{1}$, $C_{2}$,
$C_{\mathrm{num}}$, $C_{\mathrm{den}}$, which are fixed by
reproducing the bulk properties of nuclear matter~\cite{Ghim:2021odo} 
\begin{align}
    & C_{\mathrm{cl}} = -0.0561,
      \quad C_{1} = 0.6434, 
      \quad C_{2} = -0.1218,
      \quad C_{\mathrm{num}} = 65.60,
      \quad C_{\mathrm{den}}=0.60.
  \end{align}
We find that the present medium-modified EoS describes very well 
bulk properties of nuclear matter and symmetry energies up to density
$\sim 3\rho_{0}$. Even at higher densities, the results are in
reasonable agreement with empirical data.  
We want to examine the EoS developed in Ref.~\cite{Ghim:2021odo} by
computing the mass-radius relation of neutron stars in the current
work. For simplicity, we also introduce the common
strangeness mixing parameter $\chi$ as
\begin{align}
\sum_\mathrm{s}\mathrm{s} \delta_{\mathrm{s}}=\chi,
\end{align} 
where $s$ is the strangeness of the hyperon.
Then the binding energy per baryon is given as a function of three
parameters $\varepsilon=\varepsilon(\lambda,\delta,\chi)$.

The nuclear symmetry energy is defined as the second derivative of the
binding energy with respect to $\delta$
\begin{align}
\varepsilon_{\mathrm{sym}}\left(\lambda\right) 
& =  
\frac{1}{2!}
\left.\frac{\partial^{2}\varepsilon
 \left(\lambda,\delta,0,0,0\right)}{\partial\delta^{2}}
\right|_{\delta=0} .
\label{eq:symen}
\end{align}
$\varepsilon_{\mathrm{sym}}$ can be expanded in the vicinity of the
saturation point $\lambda=1$ 
\begin{align}
\varepsilon_{\mathrm{sym}}\left(\lambda\right) 
& =  
a_{\mathrm{sym}}+\frac{L_{\mathrm{sym}}}{3}\left(\lambda-1\right)
+K_{\mathrm{sym}}\frac{\left(\lambda-1\right)^{2}}{18}
+\cdots,
\label{eq:symE}
\end{align}
where $a_{\mathrm{sym}}$ denotes the nuclear symmetry energy at 
the saturation point, $L_{\mathrm{sym}}$ stands for that of its slope
parameter $L_{\mathrm{sym}}$, and $K_{\mathrm{sym}}$ represents the
asymmetric part of the incompressibility. They are expressed as 
\begin{align}
a_{\mathrm{sym}} 
& =  
-\frac{9}{20}\frac{C_{\mathrm{num}}\left(b-7r/18\right)}
                    {\left(1+C_{\mathrm{den}}\right)},\\ 
L_{\mathrm{sym}}
&=
-\frac{27}{20}\frac{C_{\mathrm{num}}\left(b-7r/18\right)}
                   {\left(1+C_{\mathrm{den}}\right)^{2}},\\ 
K_{\mathrm{sym}}&=\frac{81C_{\mathrm{den}}
                  C_{\mathrm{num}}(b-7r/18)}{10(1+C_{\mathrm{den}})^{3}},  
\label{eq:al}
\end{align}
where  $b= (m_{\mathrm{d}}-m_{\mathrm{u}})\beta$ and
$r=(m_{\mathrm{d}}-m_{\mathrm{u}})\gamma$.
The empirical values of $a_{\mathrm{sym}}$ and $L_{\mathrm{sym}}$ are
known to be $a_{\mathrm{sym}}=(31.7\pm3.2)$ MeV and
$L_{\mathrm{sym}}=(58.7\pm 28.1)$ MeV,
respectively~\cite{Oertel:2016bki}. 
In the current work, we
take four different sets of parameters for $a_{\mathrm{sym}}$ and
$L_{\mathrm{sym}}$ as shown in Table~\ref{tab:1}. 

\begin{table}[hbt]
\caption{Four different sets of the parameters for the symmetry
  energy and slope parameter at the normal nuclear matter density
  $\rho_0$.}  
\label{tab:1}
\begin{tabular}{lcccc}\hline \hline
 & $a_{\rm sym}$ [ MeV] & $L_{\rm sym}$ [MeV] & 
$C_{\rm num}$ &  $C_{\rm den}$ \\
 \hline
 Set I & 30& 50& 69.20 & 0.80 \\
 Set II & 32 & 50 & 78.72 & 0.92\\
 Set III & 30 & 60& 57.66 & 0.50 \\
 Set IV & 32 & 60 & 65.60 & 0.60\\
 \hline\hline
\end{tabular}
\end{table}
The density dependence of the symmetry energy is constrained by the
experimental data on heavy-ion collision~\cite{Tsang:2004zz,
  Liu:2006xs, Famiano:2006rb, Tsang:2008fd}. In the previous
work~\cite{Ghim:2021odo}, we showed that the results for
$\varepsilon(\lambda)$ are in good agreement with the APR
predictions~\cite{Akmal:1998cf} and IAS
constraints~\cite{Danielewicz:2013upa}. 
The values of $a_{\mathrm{sym}}$ and $L_{\mathrm{sym}}$ are also very
similar to those in various theoretical works~\cite{Lattimer:2012nd,
Ozel:2016oaf, Hornick:2018kfi, Wang:2022sev}.  
As will be shown later, the results for the masses and radii of the
neutron stars are not sensitive to changes of the symmetry energy,
whereas its slope parameter has certain effects on them. In
Ref.~\cite{Ghim:2021odo}, we also discussed as to how the parameters, 
$K_{\mathrm{sym}}$ and related incompressibilities, were determined in
detail. The results for both the energy and pressure indicate that the
slope parameter of the symmetric energy will influence the mass-radius
relations of the neutron stars.

 \begin{figure}[htp]
   \centering
   \includegraphics[scale=0.19]{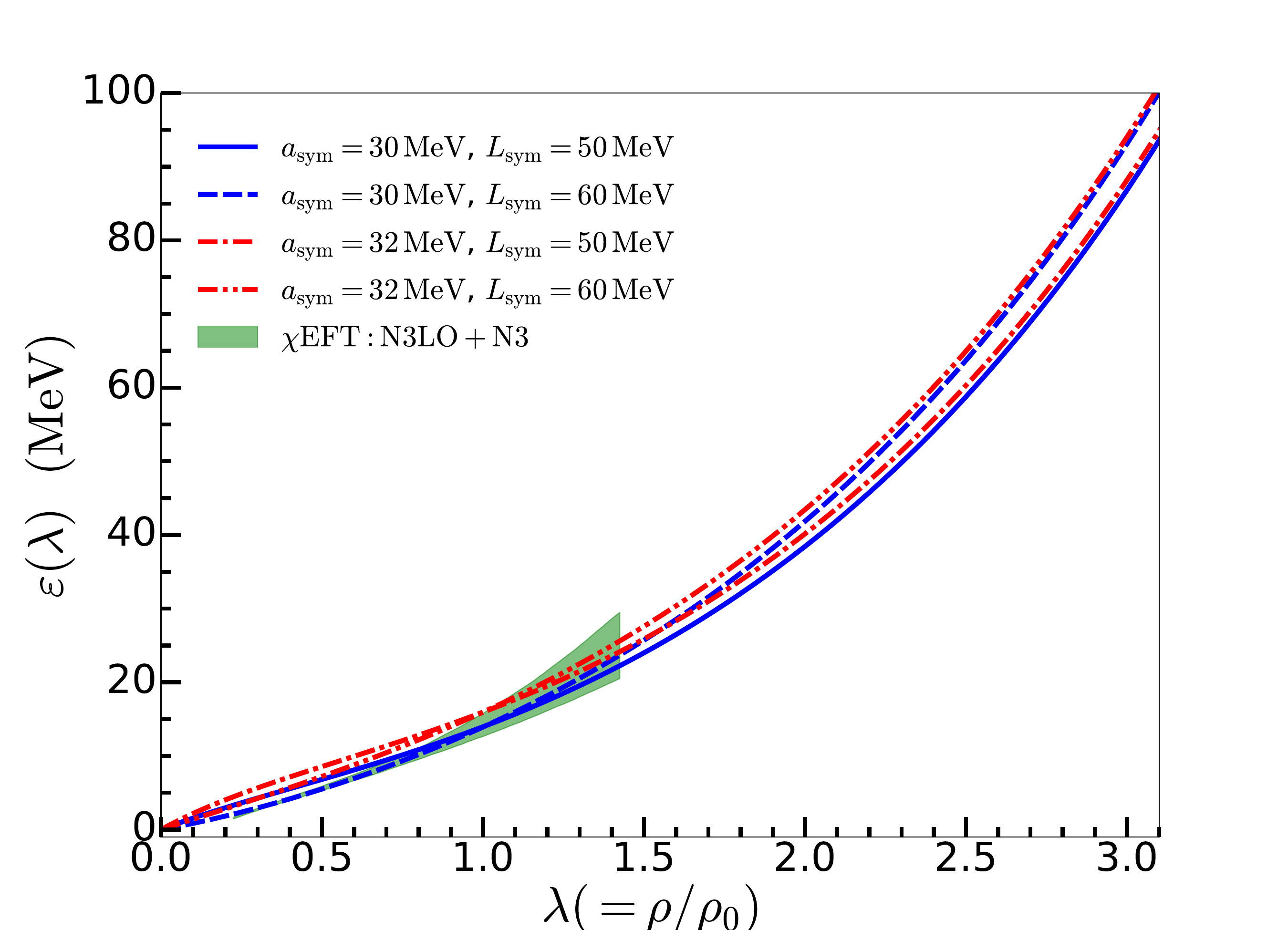}
   \includegraphics[scale=0.19]{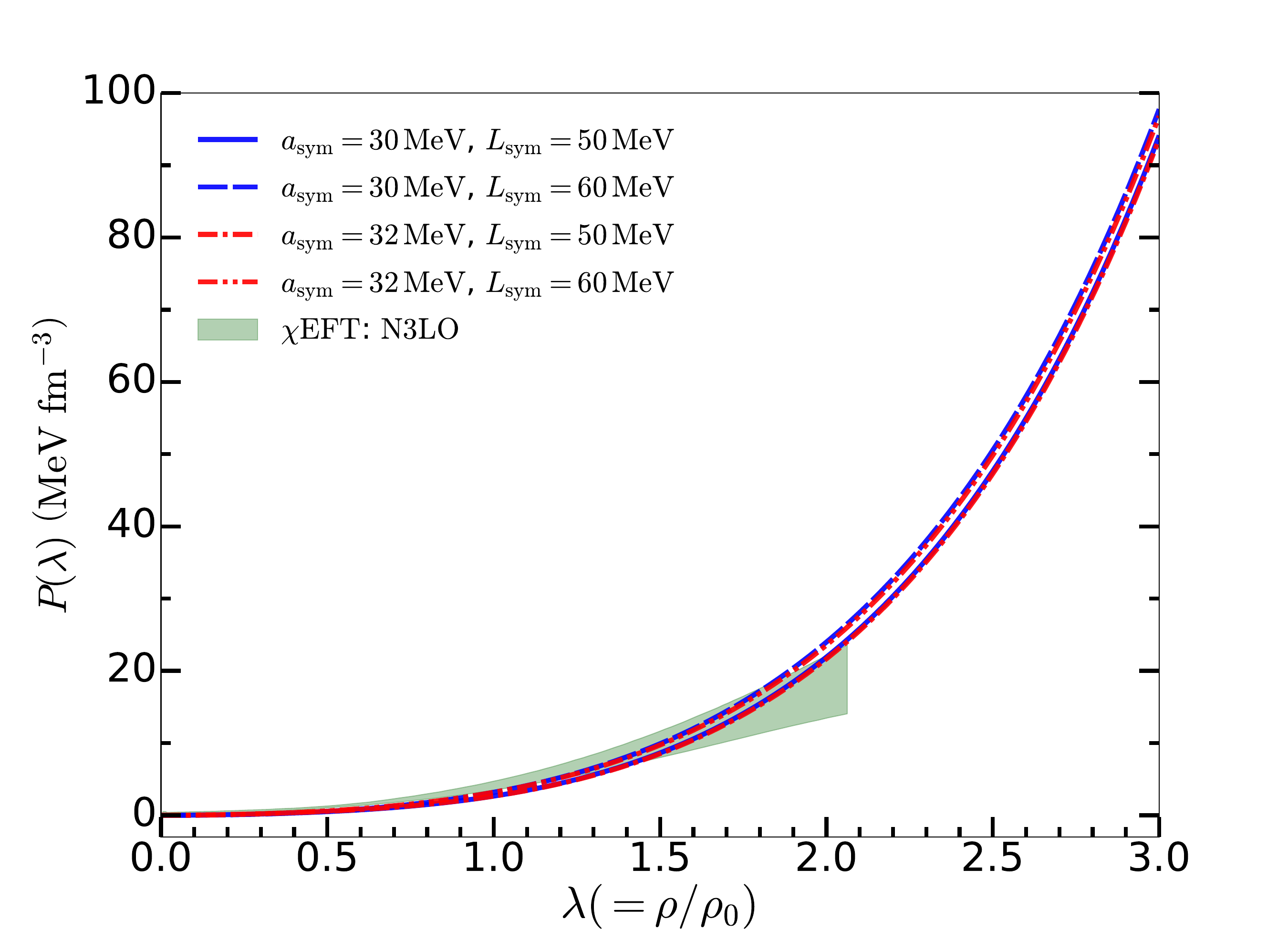}
   \caption{The energy and pressure as functions of
     $\lambda$. The parameters in the legend correspond to those given
   in Table~\ref{tab:1}. The green area represents the results from 
   Ref.~\cite{Machleidt:2024bwl}.}   
   \label{fig:1}
 \end{figure}
The left panel of Fig.~\ref{fig:1} depicts the binding energy per
nucleon for neutron matter as a function of $\lambda$, of which the
expression is given by Eq.~\eqref{eq:BE} with $\delta=1$ and
$\delta_s=0$. The current results are in good agreement with those
obtained in chiral effective theory~\cite{Machleidt:2024bwl} with the
three-loop order corrections and three-body interaction
considered. We find one interesting feature of the slope parameter
$L_{\mathrm{sym}}$. When $\lambda$ is small, the symmetry energy
governs the behavior of the energy per nucleon. As $\lambda$
increases, the slope parameter takes charge of it. 
The right panel of Fig.~\ref{fig:1} draws the pressure as a function
$\lambda$. We also compare the results with those from
Ref.~\cite{Machleidt:2024bwl}. As observed in the right panel of
Fig.~\ref{fig:1}, a larger value of the slope parameter, i.e., 
$L_{\mathrm{sym}}=32$ MeV enhances the pressure. On the other hand,
the change of the symmetry energy shows only a tiny effect on the
pressure.  

\section{Energy densities and pressures for neutron stars}
\label{subsec:NS}
To study the mass-radius relation of the neutron stars, we
consider static and spherically symmetric non-rotating neutron stars,
which satisfy the TOV equation. The mass function of the neutron star
$M(r)$ is given as a function of radius $r$ 
 \begin{align}
\mathcal{M}(r)=4\pi\int_{0}^{r} \mathrm{d}r\, r^{2}
\mathcal{E}(r),
\label{eq:LMr}
 \end{align}
where $\mathcal{E}(r)$ stands for the energy density as a
function of $r$. Then the total mass of the neutron star with radius
$R$ is defined by 
 \begin{align}
  M=\mathcal{M}(R).
\label{eq:NMM}
 \end{align} 
The pressure $P(r)$ is determined by solving the TOV equation  
\begin{align}
\frac{d P(r)}{dr}&=\frac{\mathcal{E}(r)\mathcal{M}(r)}{r^{2}}\left( 1-
  \frac{2 \mathcal{M}(r)}{r}\right)^{-1} \left(
  1+\frac{P(r)}{\mathcal{E}(r)}\right)\left( 
1+\frac{4\pi r^{2}P(r)}{\mathcal{M}(r)} \right)
\label{eq:TOV}
\end{align}
in natural units $c=G=1$. 
The boundary conditions for the mass and energy densities at the center
of the neutron star are given as   
\begin{align}
\mathcal{M}(0)=0, \quad \mathcal{E}(0)=
\mathcal{E}_{\mathrm{c}},
\end{align} 
where the $\mathcal{E}_{\mathrm{c}}$ is the central energy density of the
neutron star. The stability condition requires the local pressure to
be zero at the surface of the neutron star 
\begin{align}
P(R)=0.
\label{eq:pzero}
\end{align}
To obtain the profile of a neutron star, we need to know the
energy dependence of the pressure 
\begin{align}
P=P(\mathcal{E}),
\end{align}
which can be derived from the following equation
\begin{align}
P(\lambda,\delta,\chi)&=\rho_{0} \lambda^{2}
\frac{\partial\varepsilon(\lambda, \delta,\chi)}{\partial
\lambda},\cr
\mathcal{E}(\lambda)&=\left[\varepsilon(\lambda,\delta,\chi)+m_{N}
\right]\lambda\rho_{0}.
\end{align} 

 \begin{figure}[htp]
   \centering
   \includegraphics[scale=0.19]{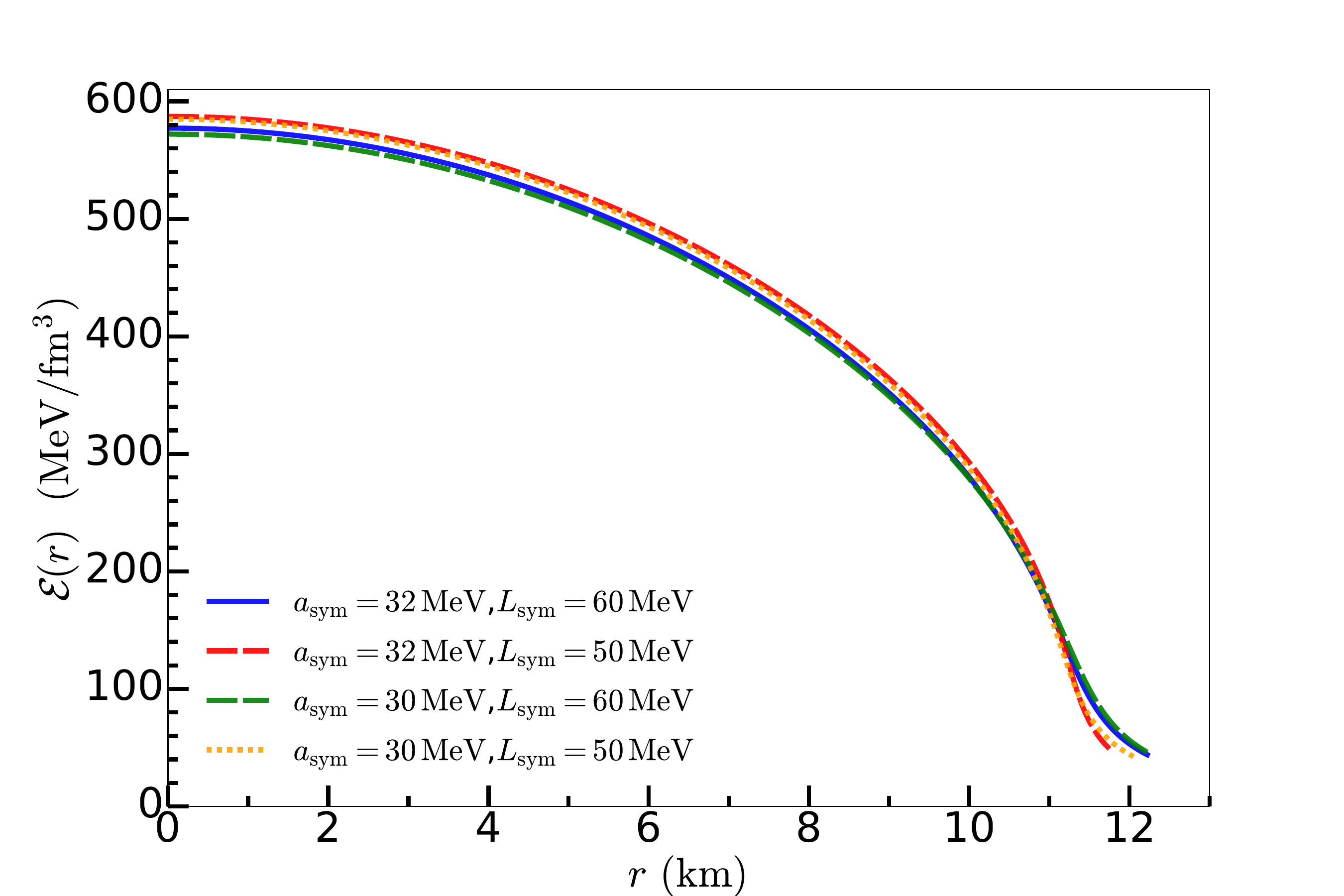}
   \includegraphics[scale=0.19]{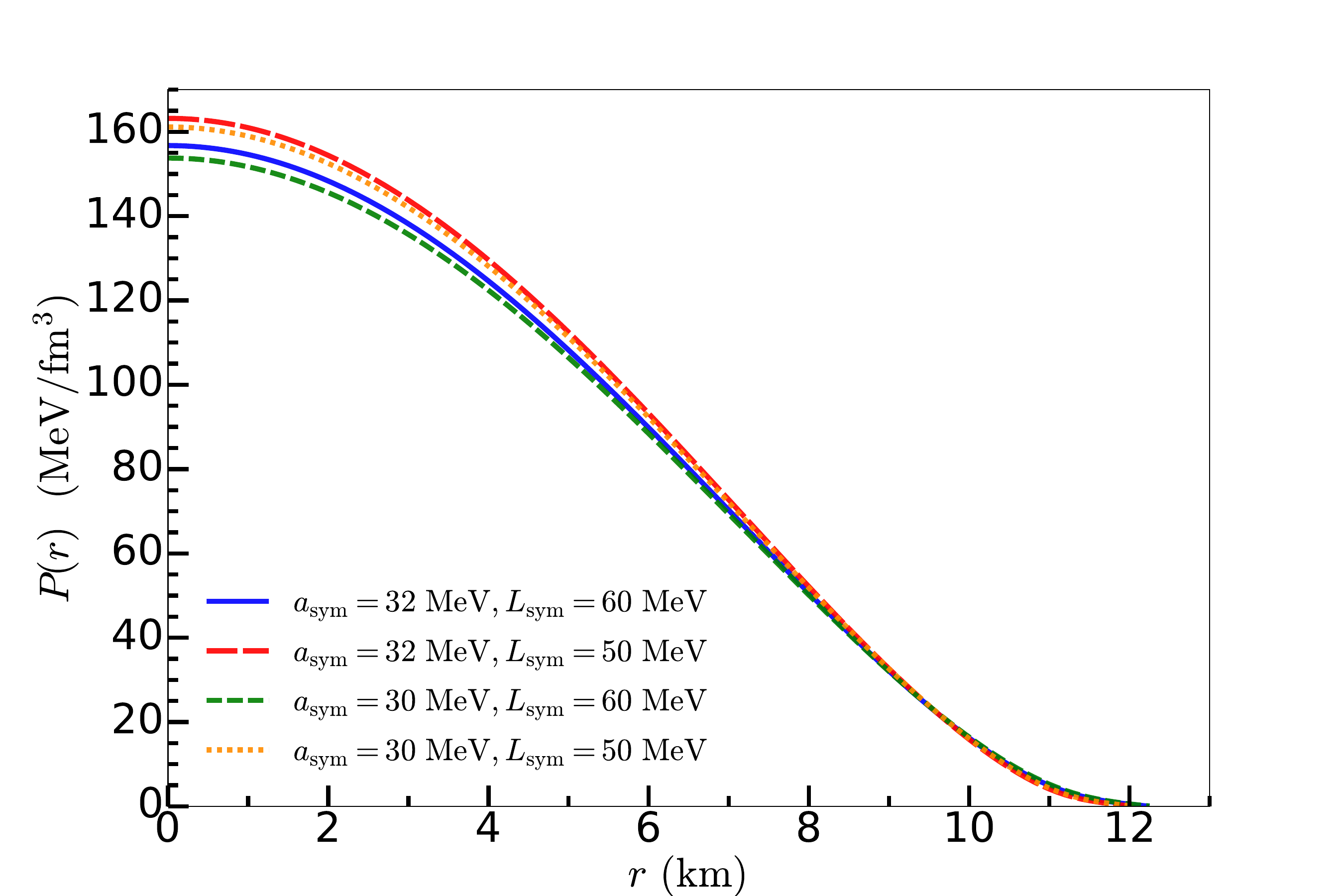}
   \caption{The energy density and pressure of the neutron star
     as functions of $r$.} 
   \label{fig:2}
 \end{figure}
The numerical result for $\mathcal{E}(r)$ is drawn in the left panel
of Fig.~\ref{fig:2} with the value of the incompressibility $K_0=240$
MeV. The results are rather stable as the parameters are changed. As
expected, the energy density falls off monotonically, as $r$
increases, and then it slowly vanishes at the surface of the neutron star. 
The right panel of Fig.~\ref{fig:2} shows the interior pressure of
the neutron star as a function of $r$, which is obtained by solving
Eq.~\eqref{eq:TOV}. The pressure also drops off as $r$ increases, and
vanishes at around 11-12 km, which corresponds to the radius of the
neutron star. Using Eq.~\eqref{eq:NMM} and \eqref{eq:pzero}, we can
derive the masses and radii of the neutron stars.  

Although the leptons inside a neutron star do not participate in
the strong interaction, this will cause certain contributions from the
protons in the crust of the neutron star. We use a relativistic
Fermi gas model to describe the electrons inside a the neutron 
star~\cite{Teukolsky, Lattimer:1985zf}, so that we obtain the energy
density and pressure of the electron gas as 
\begin{align} 
\mathcal{\epsilon}_{e} &= \frac{3}{4}n{e}\mu_{e}, \;\;\;
P_{\mathrm{e}} = \frac{1}{4}n_{e}\mu_{e},
\end{align}
where $n_{e}$ denotes the number density of the electrons, and
$\mu_{e}$ is the chemical potential of the electron. 

Since we consider the beta-stable condition, the chemical potential of
the electron can be expressed in terms of the symmetry energy
as~\cite{Steiner:2004fi}   
\begin{align}
\mu_{e} = 4 \delta E_{\mathrm{sym}}(\lambda).
\end{align}
Since the neutron star satisfies the charge neutrality condition,
the electron number density must be the same as the proton number
density, which brings about the contribution from the protons. Thus, 
the symmetry energy depends not only on the density of the neutron star 
but also on the fraction of the protons. We fix the electron
and proton fractions to be $n_{e} = n_{p} = 0.1$ in the current work,
assuming that the leptonic effects mainly come from the crust of the
neutron star. So, the electron affects only the subnuclear 
density region, i.e., $\lambda < 1$~\cite{Steiner:2004fi}.

\section{Results and discussions}
\label{sec:4} 
Before we proceed to discuss the results, we want to emphasize that
the current theoretical framework is based on linear density
approximation. Moreover, we do not include quark matter in the current
work. It implies that we can only describe physical
observables in medium approximately up to $\rho \approx 4\rho_0$. 
If the medium density becomes larger than $4\rho_0$, the speed of
sound would break the causality. Thus, in the forthcoming discussions,
we will restrict ourselves to lower densities than $4\rho_0$. 

\subsection{Neutron stars}
Since all the parameters were determined in the previous
work~\cite{Ghim:2021odo}, we do not change any parameters for
evaluating the masses and radii of the neutron stars. Using the four
different sets of parameters listed in Table~\ref{tab:1}, we proceed
to compute the mass-radius relations of the neutron stars. 

\begin{figure}[htp]
  \includegraphics[scale=0.275]{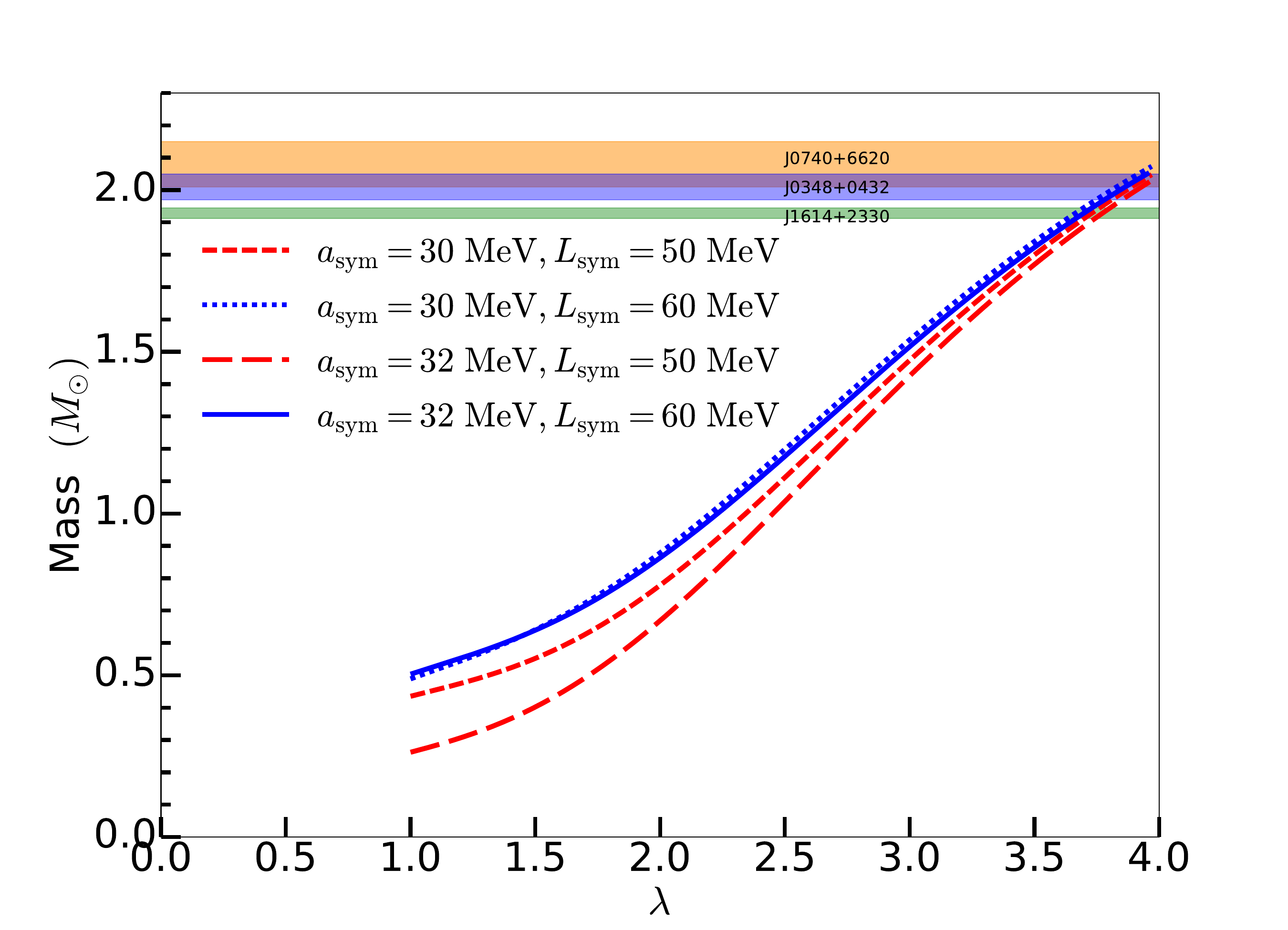} 
  \caption{Masses of the neutron star are drawn as functions of the
    dimensionless central density $\lambda=\rho/\rho_0$ with
    $\rho_0=0.16\,\mathrm{fm}^{-3}$ in unit of the solar mass
    $M_{\odot}$. The data for the orange, blue,
    and green bars are taken from
Refs.~\cite{Fonseca:2021wxt, Antoniadis:2013pzd, Fonseca:2016tux}.}    
  \label{fig:3}
\end{figure} 
In Fig.~\ref{fig:3}, we show the masses of the neutron star in units
of the solar mass as functions of dimensionless central density
$\lambda$ with the four different parameter sets used. While the
results are not sensitive to $a_{\mathrm{sym}}$, they 
depend on $L_{\mathrm{sym}}$ in lower nuclear matter density. The
results indicate that the central density of the neutron stars
observed in Refs.~\cite{Fonseca:2021wxt, Antoniadis:2013pzd,
  Fonseca:2016tux} is approximately $\lambda\approx 4$ or $\rho\approx
0.64\,\mathrm{fm}^{-3}$. However, when heavy neutron stars are
considered, quark matter must come into a significant role, which we did
not include in the current work. This, we present the results for the
masses of neutron stars up to $\lambda \approx 4$. 

It is interesting to compare the current results with those from other
approaches. In Ref.~\cite{Akmal:1998cf}, Akmal et al. constructed the
EoS by using the Argonne nucleon-nucleon potential (A18), Urbana 
three-nucleon interactions, the boost interaction, and 
the estimated maximum masses of $M = 2.2M_{\odot}$ at $\lambda\approx 6$.
Baym et al.~\cite{Baym:2017whm} also obtained the maximum masses of $M
= 2.2M_{\odot}$ saturated at around $\lambda\approx 6$. On the other
hand, Brandes et al.~\cite{Brandes:2023hma} incorporated the data on
the newly observed neutron called the black-widow (PSR
J0952-0607)~\cite{Romani:2022jhd}, and estimated $\lambda \approx 4$
for $M=2.1\,M_{\odot}$. Vijayan et al.~\cite{Vijayan:2023qrt} reported
very interesting results. Using two different EoS and considering the
medium modification of the pion mass, they found that the central
density reaches $\lambda\approx 6$ for $M=(1.7-2.3)M_{\odot}$. Since
the current framework has naturally taken into account the medium
modifications of hadrons, the present results are in good agreement
with those from Ref.~\cite{Romani:2022jhd}.

\begin{figure}[htp]
  \includegraphics[scale=0.3]{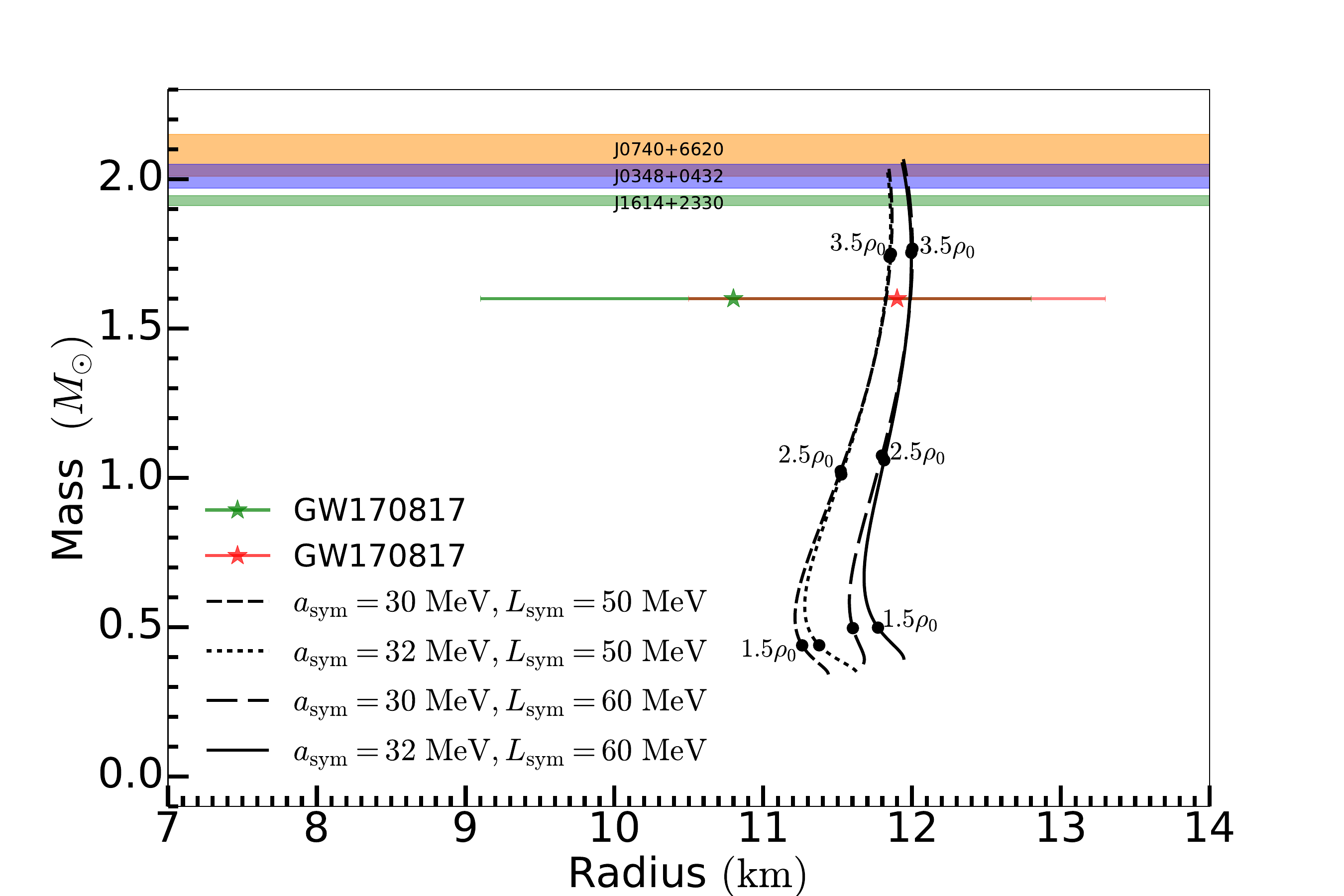} 
  \caption{Mass-radius relations of neutron stars. 
The black curves represent the results with the four different sets of
parameters given in Table~\ref{tab:1}. The value of the 
incompressibility is chosen as $K_0=240$ MeV. The red and green lines
with stars are taken from the LIGO experimental
data~\cite{LIGOScientific:2018cki}. 
The data for the orange, blue, and 
green bars are taken from
Ref.~\cite{Fonseca:2021wxt,Antoniadis:2013pzd,Fonseca:2016tux}.
}
  \label{fig:4}
\end{figure}
Figure~\ref{fig:4} depicts the mass-radius relations for the neutron
stars. As already discussed above, the mass-radius relations are not 
sensitive to $a_{\mathrm{sym}}$, the radius tends to increase as the
slope parameter $L_{\mathrm{sym}}$ increases, in particular, when the
neutron matter density is low. As the density increases, the
dependence of the mass-radius relations on $L_{\mathrm{sym}}$ becomes
smaller. Figure~\ref{fig:4} also indicates that the central densities 
in PSR J0740+6620, PSR J0348+0432, and PSR
J1614+2330~\cite{Fonseca:2021wxt, Antoniadis:2013pzd,
  Fonseca:2016tux} are estimated to be around $3.5\rho_0$.
As mentioned previously, by including the leptonic effects, the
radius of the neutron stars tend to decrease. They mainly contribute
to lower-density regions.

\begin{figure}[htp]
  \includegraphics[scale=0.3]{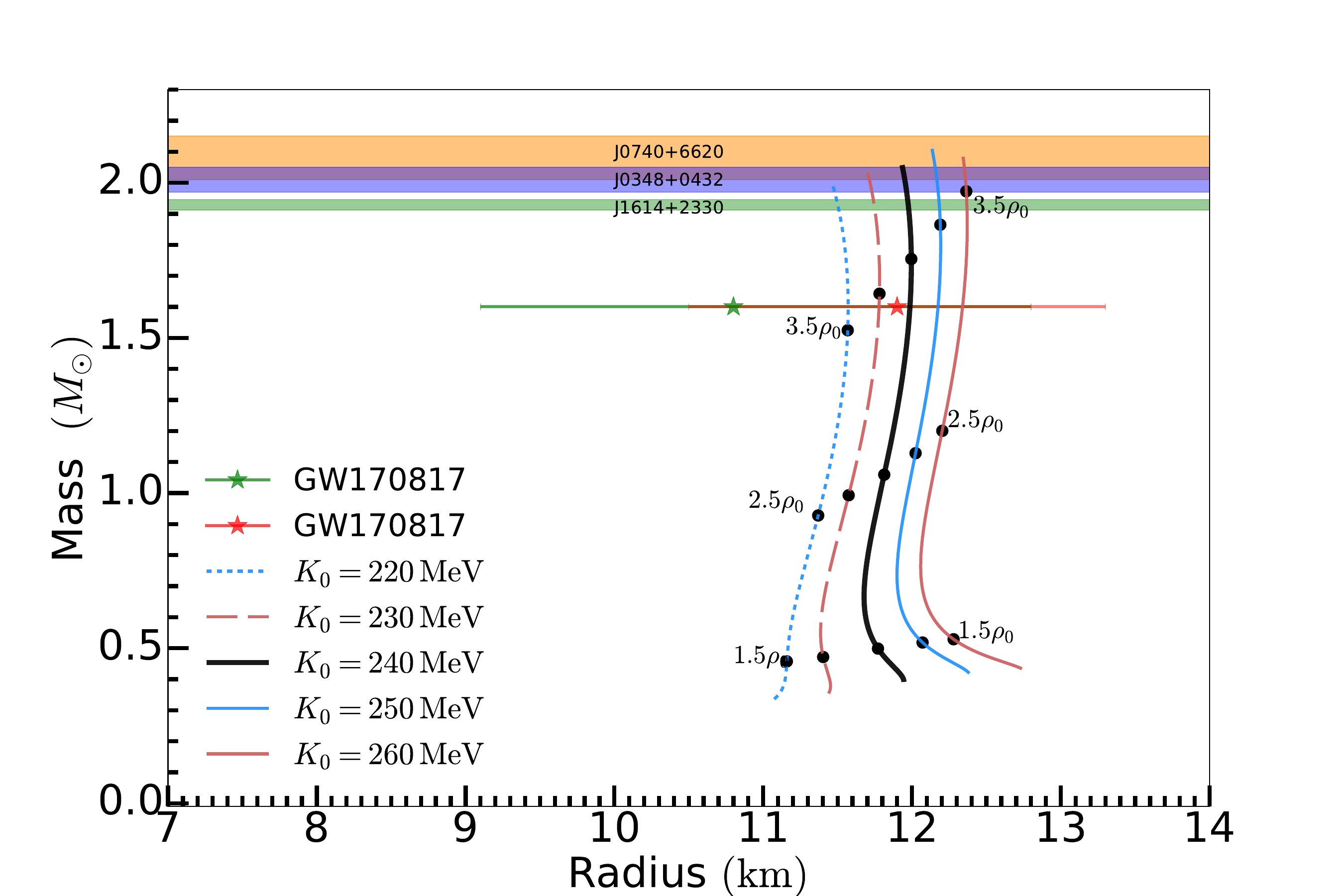} 
\caption{Dependence of the neutron star mass-radius relation on the
  incompressibility, $K_0$. The values of the symmetry energy at
  saturation density and the corresponding slope
  parameter are chosen 
  as $a_{\rm sym}=32$ MeV and $L_{\rm sym}=60$ MeV, respectively. The 
  observational and empirical data are the same as in
  Fig.\,\ref{fig:4}.}  
  \label{fig:5}
\end{figure}
The empirical values of the nuclear matter incompressibility $K_0$ 
are given in a wide range. For example, the value of 
$K_0$ was often predicted as $K_0\approx 
(290\pm70)$\,MeV~\cite{Sharma:1988zza, Shlomo:1993zz, 
  Ma:1997zzb, Vretenar:2003qm, TerHaar:1986xpv, Brockmann:1990cn}. 
In Ref.~\cite{Piekarewicz:2003br}, however, a smaller value was
extracted from the data on the isoscalar giant monopole resonance in
${}^{208}\mathrm{Pb}$, based on relativistic mean-field models:
$K_0=(248\pm 8)$ MeV. In Ref.~\cite{Stone:2014wza}, $K_0= (240\pm  
20)$\,MeV was extracted from a reanalysis of the data on the energies
of the giant monopole resonances in even-even ${}^{112-124}$Sn
and ${}^{106,100-116}$Cd and earlier data on $58\le A \le 208$
nuclei. A similar value was also estimated in
Ref.~\cite{Shlomo:2006ole}.  Thus, while we choose 
$K_{0}=240\,\mathrm{MeV}$, it is of great interest to examine the
dependence of the mass-radius relations of the neutron stars on
$K_0$. In Fig.~\ref{fig:5}, we demonstrate the mass-radius relations
of the neutron stars, varying the incompressibility of nuclear matter
$K_0$ from 210 MeV to 260 MeV. We take the values of
$a_{\mathrm{sym}}=32$ MeV and $L_{\mathrm{sym}}=60$ MeV. The tendency
of the mass-radius relations is obvious in general. If we increase 
$K_0$ from 220 MeV to 260 MeV, the mass-radius relations 
remain as almost the same shape but are shifted to the right. So, the
radii of the neutron stars get larger as $K_0$ increases. 

In Table~\ref{tab:2}, we list the results for the central densities
and the radii, which correspond to two different values of the neutron
star masses, i.e. $M=1.4M_{\odot}$ and $M=2.0M_{\odot}$. 
We take different values of $K_0$ with two different sets of
parameters for neutron matter. The results show that as $K_0$
increases, the central density decreases, so the radius of the neutron
star enlarges.  
\begin{table}[htp]
\caption{central densities and radii of the neutron stars with two
  different values of the mass. We take three different values of the
  incompressibility of nuclear matter with the two different sets of
  symmetry energy and slope parameters at  saturation density
  $\rho_0$.  $n$ denotes the central density of neutron matter,
  $\lambda$ stands for the corresponding dimensionless central 
density, and $R$ designates the radius of the neutron star. Note that
$n_0$ corresponds to $\rho_{0}=2.677\times 10^{14}$ g/cm$^{3}$.}   
  \label{tab:2}
    \begin{center}
    \begin{tabular}{c|ccc|ccc}
      \hline \hline
      & 
      & 1.4$M_{\odot}$ & &
      & 2.0$M_{\odot}$ & \\
    $K_{0}$, $a_{\mathrm{sym}}$, $L_{\mathrm{sym}}$ &
    $n$  & $\lambda$  & $R$ &
    $n$  & $\lambda$  & $R$ \\
    $[ \mathrm{MeV} ]$ & 
    $[\mathrm{fm}^{-3}]$ &[dimensionless] & [km]&
    $[\mathrm{fm}^{-3}]$ &[dimensionless] & [km]\\
    \hline \hline
210 MeV, 30 MeV, 50 MeV &
0.502& 3.136& 11.514&
0.672& 4.198& 11.308\\
240 MeV, 30 MeV, 50 MeV &
0.429& 2.681& 12.121&
0.549& 3.430& 12.094\\
270 MeV, 30 MeV, 50 MeV &
 0.385& 2.406& 12.542&
 0.477& 2.983& 12.642\\
\hline
210 MeV, 32 MeV, 50 MeV &
0.485& 3.030& 11.490&
0.646& 4.036& 11.281\\
{240 MeV, 32 MeV, 50 MeV} &
 0.454&  2.839&  12.110&
 0.587&  3.666&  12.074\\
270 MeV, 32 MeV, 50 MeV &
0.434& 2.712& 12.547&
0.550& 3.435& 12.629\\
\hline
210 MeV, 30 MeV, 60 MeV &
0.480& 2.997& 11.858&
0.658& 4.115& 11.513\\
240 MeV, 30 MeV, 60 MeV &
0.413& 2.584& 12.435&
0.538& 3.363& 12.299\\
270 MeV, 30 MeV, 60 MeV &
0.372& 2.328& 12.818&
0.468& 2.927& 12.834\\
\hline
210 MeV, 32 MeV, 60 MeV &
0.485& 3.032& 11.826&
0.663& 4.146& 11.473\\
240 MeV, 32 MeV, 60 MeV &
0.417& 2.609& 12.426&
0.541& 3.384& 12.272\\
270 MeV, 32 MeV, 60 MeV &
0.375& 2.347& 12.831&
0.470& 2.944& 12.820\\
\hline \hline
    \end{tabular}
    \end{center}
\end{table}
\subsection{Strageness-mixed stars}
A virtue of the current approach is that we can easily incorporate the
hyperons into nuclear matter. As shown in Eq.~\eqref{eq:delta,lambda},
the parameter $\delta_s$ controls the proportionality of the
hyperons with strangeness $s=|S|$. Thus, the energy density can depend
on $\delta_s$ as in Eq.~\eqref{eq:BE}. Once we construct the EoS for
baryonic matter with strangeness, we can evaluate the mass-radius
relations of hyperon stars. Before we compute them, we will
recapitulate how baryonic matter can be formulated within the present
approach. Note that we will not introduce any new parameters but
introduce the hyperons with strangeness $s=|S|$. 

We first expand the binding energy per baryon given in
Eq.(\ref{eq:BE}) with respect to isospin-asymmetry parameter $\delta$
and strangeness-mixing parameters $\delta_s$
\begin{align}
\varepsilon(\lambda,\delta,\delta_{1},\dots)  &=  
\varepsilon_{V}\left(\lambda\right)
+\varepsilon_{\mathrm{sym}}\left(\lambda\right)\delta^{2}
+\sum_{s=1}^3
\left.\frac{\partial\,\varepsilon 
\left(\lambda,\,\delta,\,\delta_{1},\dots\right)}{
\partial\,\delta_{s}}\right|_{\delta =\delta_{1}=\dots=0}\delta_{s}
                                                \cr 
&+
\frac{1}{2}\sum_{s,p=1}^3\left.\frac{\partial^{2}\, 
\varepsilon\left(\lambda,\,\delta,\,\delta_{1},\dots\right)}{
\partial\,\delta_{s}\partial\delta_{p}}\right|_{\delta=\delta_{1}=\dots=0} 
\delta_{s}\delta_{p} +\cdots,
\label{eq:Eddi}
\end{align}
where the first and second terms denote the 
volume and symmetry energies for ordinary nuclear  
matter. It is obvious that the linear terms in $\delta$
are absent due to the quadratic dependence of the binding energy per
baryon on it.  Assuming that higher-order terms in $\delta_s$ are
negligible, we choose $f_s$'s similar to $f_0$ given in
Eq.~\eqref{eq:f0}. We will parametrize $f_s$ that
is independent of $\delta$, so that we keep the nonstrange sector 
intact. For detailed analysis, we refer to Ref.~\cite{Ghim:2021odo}.

\begin{figure}[htp]
   \centering
   \includegraphics[scale=0.19]{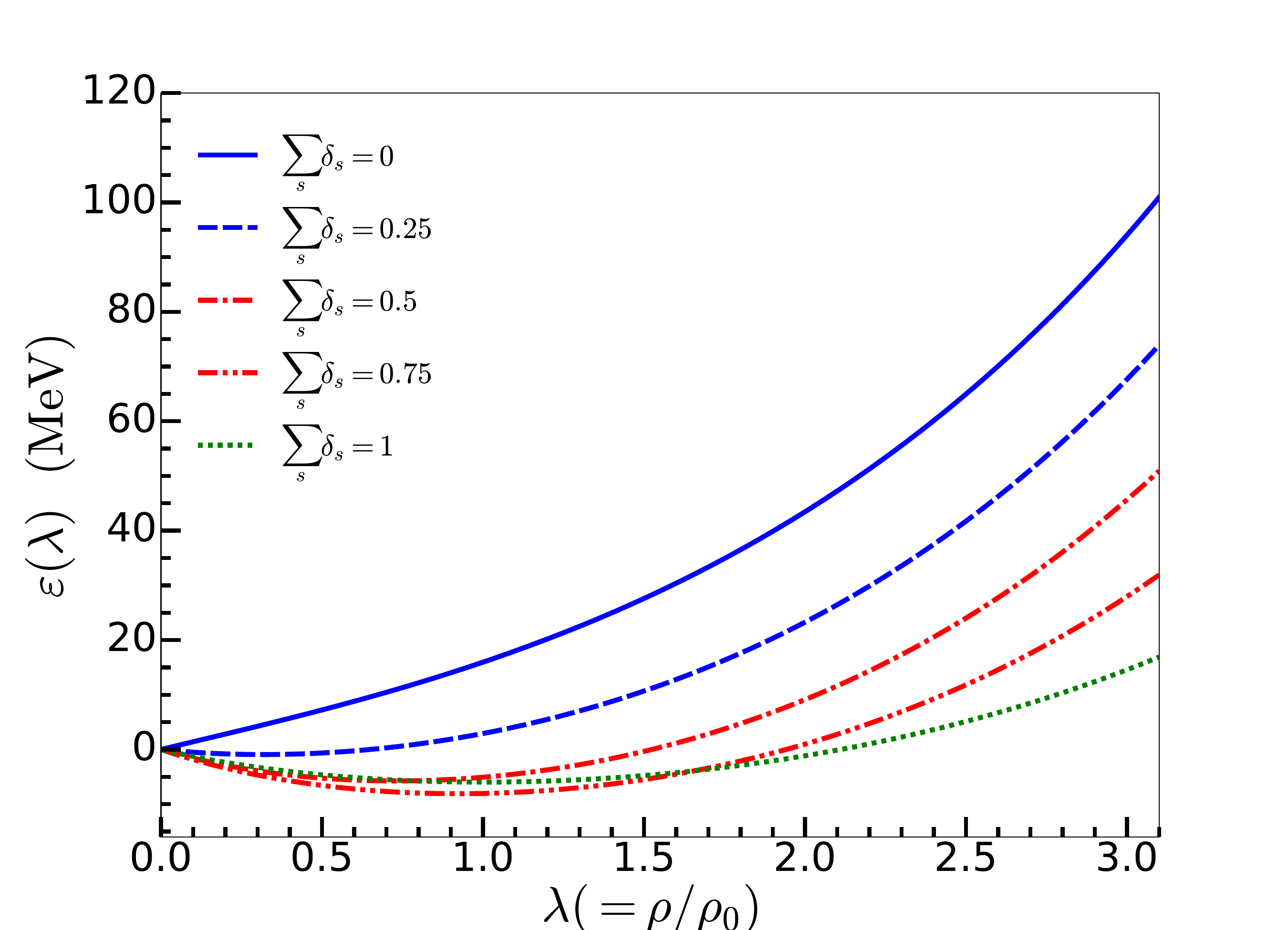}
   \includegraphics[scale=0.19]{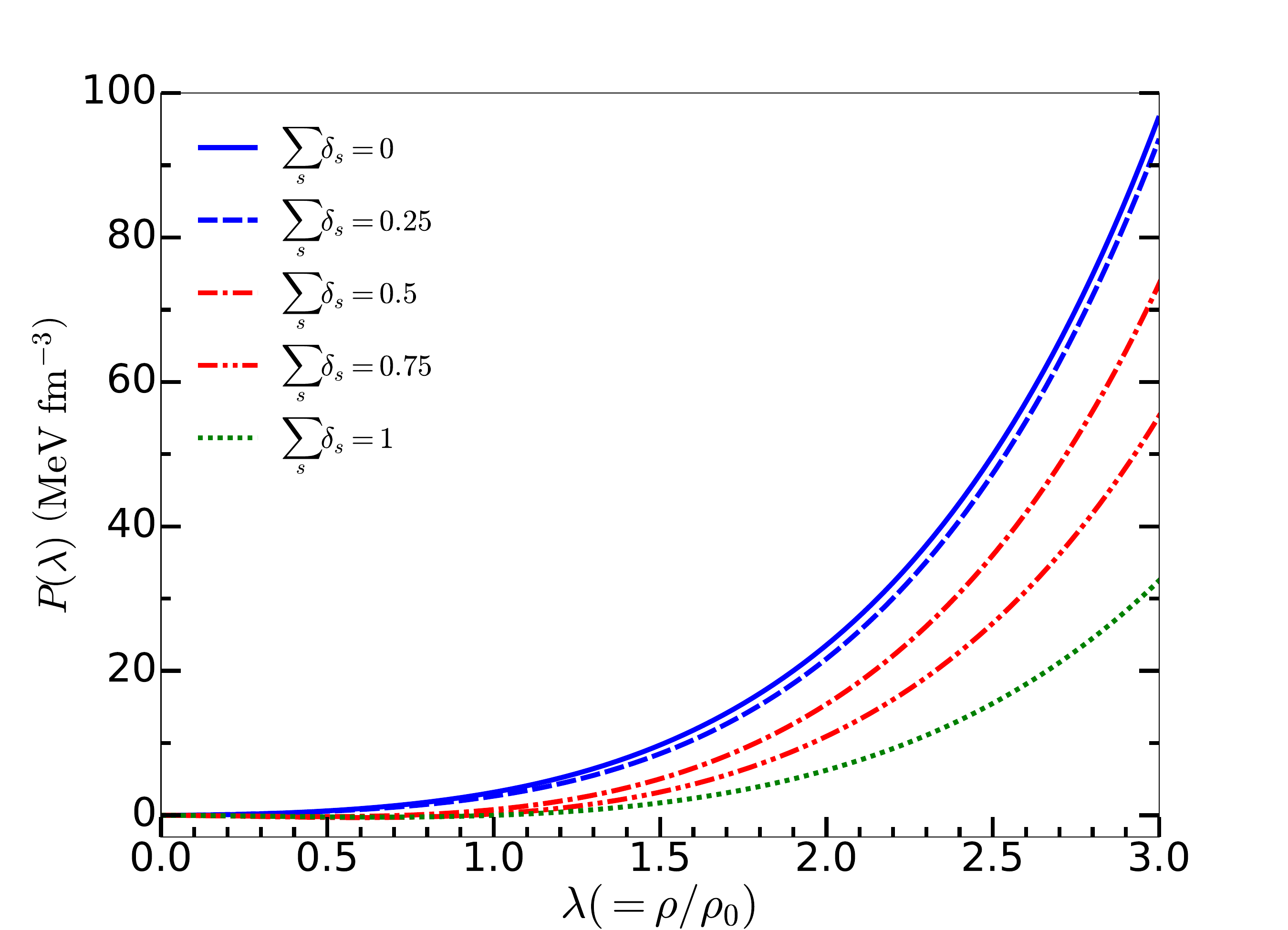}
   \caption{The energy and pressure for baryonic matter as functions
     of $\lambda$. The sum of the parameters $\delta_s$ varies from 0
     to 1.}   
   \label{fig:6}
 \end{figure}
In Fig.~\ref{fig:6}, we draw the results for the energy and pressure
as functions of $\lambda$ with the sum of the parameters $\delta_s$
changed from 0 to 1. While $\sum \delta_s = 0$ corresponds to pure
neutron matter, $\sum \delta_s=1$ indicates pure strange matter.
The left panel of Fig.~\ref{fig:6} shows that as the
number of the strange quarks increases, the energy per nucleon starts
to decrease.  When $\lambda$ reaches about $0.16$,
$\varepsilon(\lambda)$ becomes negative and turns positive, which
exhibits a behavior similar to normal nuclear matter.
As the content of the strangeness increases, the saturation point gets
lower till it reaches $-5.97$ MeV at $\lambda\simeq 1$ for pure
strange matter. However, strangeness-mixed matter has a shallow 
saturation point in comparison with nuclear
matter~\cite{Ghim:2021odo}. Consequently, the 
pressure of strangeness-mixed matter becomes weaker as the content of 
the strangeness increases, as demonstrated in the right panel of
Fig.~\ref{fig:6}. Interestingly, the pressure is lessened by about
four times for pure strange matter. This implies that the EoS for
strangeness-mixed matter is softer than neutron matter, as discussed
in Refs.~\cite{Baldo:1999rq,Vidana:2000ew,Nishizaki:2001in}.      

\begin{figure}[htp]
  \includegraphics[scale=0.3]{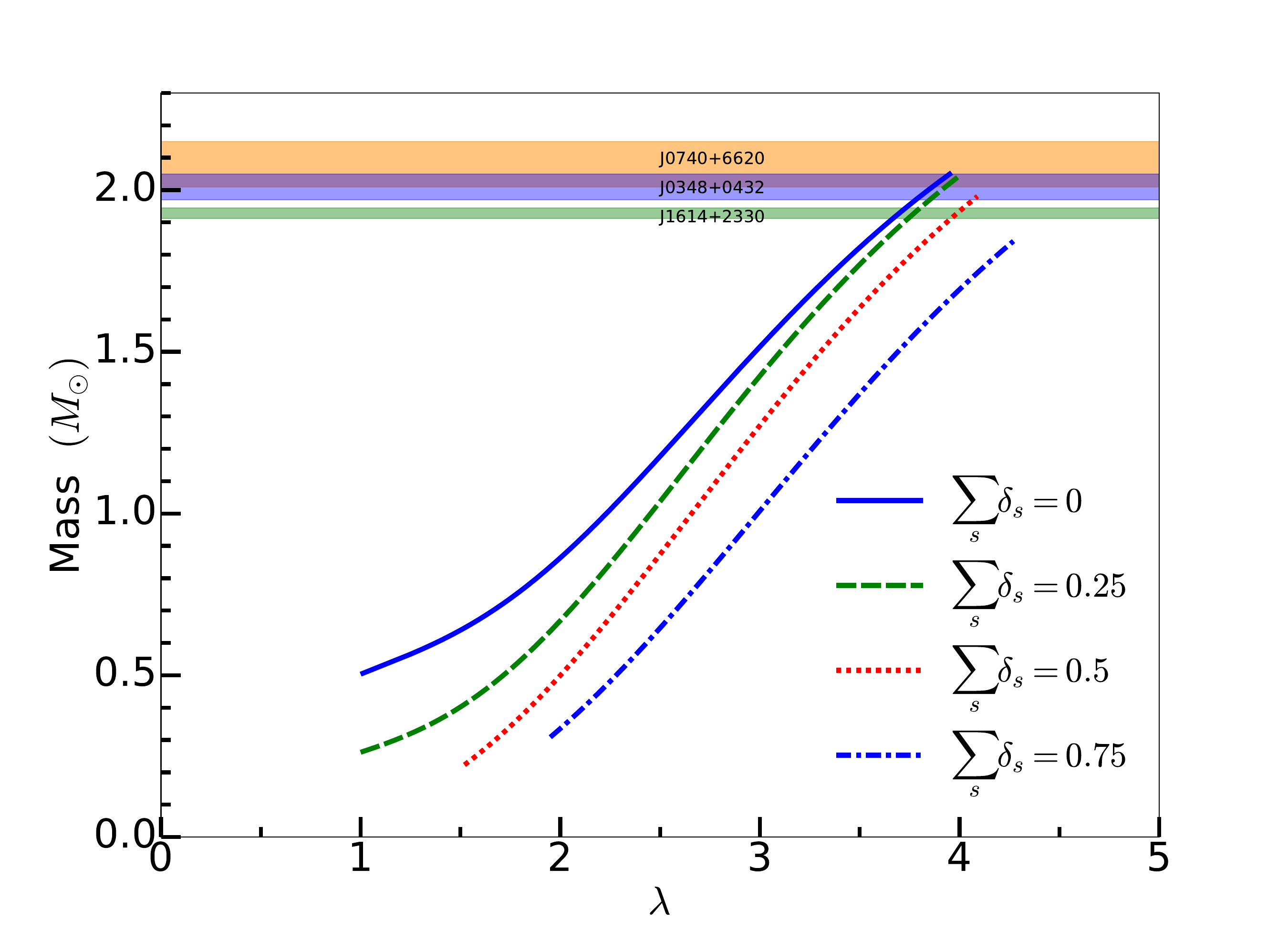} 
  \caption{Masses of the strangeness-mixed stars in unit of the solar
    mass as functions of $\lambda$ with the strangeness content varied
    from zero to 1. The values of the symmetry energy, the slope
    parameter, and the incompressibility at the saturation density are
    taken as $a_{\mathrm{sym}}=32\, \mathrm{MeV}$, $L_{\mathrm{sym}}=60\,
    \mathrm{MeV}$, and $K_{0}=240$\,MeV. The observational and
    empirical data are the same as in Fig.\,\ref{fig:4}.
  }
  \label{fig:7}
\end{figure}
In Fig.~\ref{fig:7}, we draw the results for the masses
of the strangeness-mixed stars as functions of $\lambda$. The short
dashed curve depicts the results for the neutron star mass with the
parameters $a_{\mathrm{sym}}=32\, \mathrm{MeV}$,
$L_{\mathrm{sym}}=60\, \mathrm{MeV}$, and $K_{0}=240$\,MeV used.
If we add the strange quarks by 25~\%, the mass is not much changed
except for the lower density region ($\lambda \ge 3$), as shown by
the dashed curve. If we increase the strange quark content,
the masses of the strangeness-mixed star become smaller at the same
central density. To get the masses of the known neutron stars with
higher strangeness content, the central densities should become
larger. It implies that as the EoS gets softer, the central density
becomes larger. Similar conclusions are found in
Refs.~\cite{Baldo:1999rq,Vidana:2000ew,Nishizaki:2001in,Nishizaki:2002ih,
  Chen:2011my, Masuda:2012kf}. 

\begin{figure}[htp]
  \includegraphics[scale=0.3]{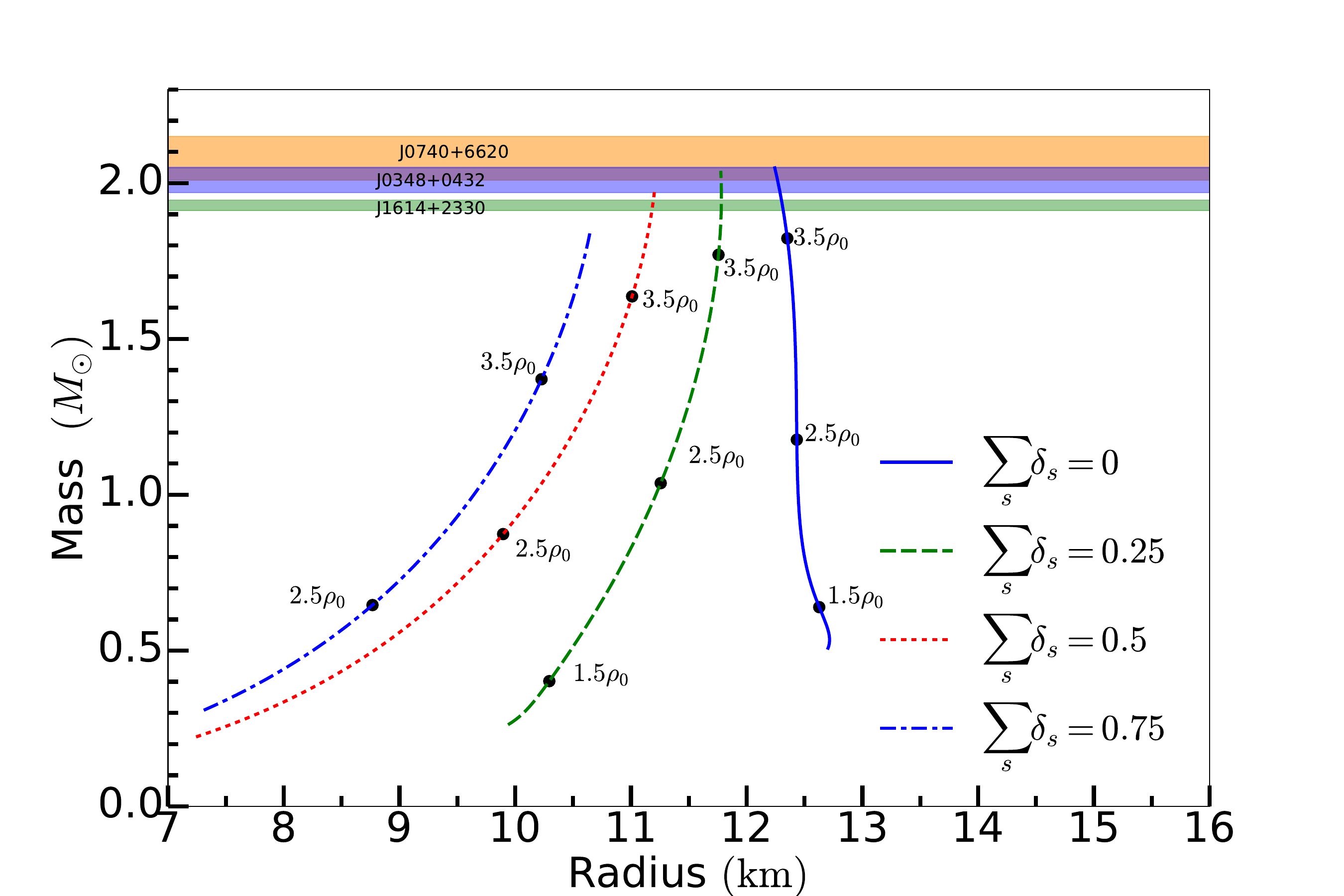} 
\caption{The mass-radius relation of a hyperon-mixed star that
has the different portion of hyperons.  The parameters of EOS are
fixed at values $K_{0}=240$\,MeV,  $a_{\mathrm{sym}}=32\,
\mathrm{MeV}$ and $L_{\mathrm{sym}}=60\, \mathrm{MeV}$. The portion of
protons is equal to zero. The results from the experiment and
analyses are given in Fig.\,\ref{fig:4}.} 
  \label{fig:8}
\end{figure}
The numerical results for the mass-radius relations of the
strangeness-mixed stars are depicted in Fig.~\ref{fig:8}. We fix 
the parameters for strangeness-mixed matter as $K_0=240$ MeV,
$a_{\mathrm{sym}}=32$ MeV, and $L_{\mathrm{sym}}=60$ MeV, which are
considered as our final set of the parameters. We compare the current
results with the observed data for the neutron stars. 
The solid curve designates the result for the neutron stars. When we
add the hyperons or the strange quarks to neutron matter, the EoS gets
softer, as discussed previously. Consequently, the mass-radius
relations become very different from those of the neutron stars. The
most prominent feature is that the central density starts to increase,
as the strangeness content increases. Assuming that the recently found
neutron stars PSR J0740+6620, PSR J0348+0432, and PSR
J1614+2330~\cite{Fonseca:2021wxt,   Antoniadis:2013pzd,
  Fonseca:2016tux} to be strangeness-mixed stars, the corresponding
central densities become larger as the strangeness content increases. 
It also indicates that the strangeness-mixed stars become more compact
than the neutron stars, which is understandable. In particular, when
the central density gets lower, the mass-radius relations of the
strangeness-mixed stars exhibit the propensity to be more shifted to
the left.  
In Table~\ref{tab:3}, varying the strangeness content from 0 to 0.25,
we list the corresponding results for the central densities 
and radii of the strangeness-mixed stars with two different masses
$M=1.4M_{\odot}$ and $M=2.0M_{\odot}$ fixed. 
\begin{table}[htp]
\caption{The central densities and radii of the strangeness-mixed
  stars with two different masses, $M=1.4M_{\odot}$ and
  $M=2.0M_{\odot}$, fixed. $n$ denotes the central number density,
  $\lambda$ is the corresponding ratio $n/n_0$, and $R$ stands for the
  radius of the strangeness-mixed star. We vary  the strangeness
  content $\delta_s$ from 0 to 0.25.}
  \label{tab:3}
    \begin{center}
    \begin{tabular}{l|ccc|ccc}
    \hline \hline
    &
    & 1.4$M_{\odot}$
    &
    &
    & 2.0$M_{\odot}$
    & \\
    \quad$ \delta_{\mathrm{s}}$ 
& $n$  
& $\lambda$  
& $R$ 
& $n$  
& $\lambda$  
& $R$ 
\\
& $[\mathrm{fm}^{-3}]$ 
& [dimensionless]
& [km]
& $[\mathrm{fm}^{-3}]$ 
& [dimensionless]
& [km]
\\
\hline \hline 
$\delta_{\mathrm{s}}=0$
& 0.417
& 2.609
& 12.426
& 0.541
& 3.384
& 12.272
\\
$\delta_{\mathrm{s}}=0.05$ 
& 0.423
& 2.641
& 12.239
& 0.544
& 3.402
& 12.173
\\
$\delta_{\mathrm{s}}=0.10$
& 0.428
& 2.672
& 12.064
& 0.548
& 3.424
& 12.075
\\
$\delta_{\mathrm{s}}=0.15$
& 0.433
& 2.706
& 11.896
& 0.552
& 3.450
& 11.977
\\
$\delta_{\mathrm{s}}=0.20$ 
& 0.439
& 2.743
& 11.736
& 0.557
& 3.480
& 11.879
\\
$\delta_{\mathrm{s}}=0.25$
& 0.445
& 2.782
& 11.580
& 0.562
& 3.514
& 11.780
\\
\hline \hline
\end{tabular}
\end{center}
\end{table}
    
\section{Summary and outlook}
\label{sec:5}
We have investigated the properties of the neutron stars and
strangeness-mixed stars, using the equation of the state for both
neutron matter and strangeness-mixed matter, which were derived from 
the medium-modified pion mean-field approach (chiral quark-soliton
model)~\cite{Ghim:2021odo}. All the relevant parameters were already
fixed in Ref.~\cite{Ghim:2021odo}. A great virtue of this approach is
that the medium modifications of the nucleon and hyperons have been
considered, in addition to describing the bulk properties of nuclear
media. In addition, we have considered the leptonic effects, which
mainly come from the crust of the neutrons stars. 
The results for the mass-radius relations were well described
in the current work. The central densities in PSR J0740+6620, PSR
J0348+0432, and PSR J1614+2330~\cite{Fonseca:2021wxt,
  Antoniadis:2013pzd, Fonseca:2016tux} were obtained to be around
$3.5\rho_0$.

We also examined the dependence of the mass-radius
relations on the incompressibility $K_0$. As $K_0$ increases, central
density becomes lower and the radius increases. To investigate the
strangeness-mixed stars, we first examined the equation of state for
strangeness-mixed matter, as the strangeness content varied from 0 to
1, which correspond to neutron matter and pure strange matter,
respectively. At $\lambda\approx 0.16$, the binding energy becomes
negative and turns positive, which behaves as normal nuclear matter.
As the content of the strangeness increases, the binding energy 
becomes further lower, though strange matter has a shallow 
saturation point in comparison with nuclear
matter. As a result, the pressure of strange-mixed matter becomes
lessened as the strangeness content increases. Note that the pressure
is increased by about four times for pure strange matter. This implies
that the EoS for strangeness-mixed matter is softened. Thus, 
the masses of the strangeness-mixed star become smaller at the same
central density. We conclude that the strange stars are more compact
than the neutron stars. 

\section*{Acknowledgments}
The authors are grateful to Myung-Ki Cheoun for fruitful discussion
and valuable comments related to leptonic effects on neutron stars. 
The present work was supported by Basic Science Research Program
through the National Research Foundation (NRF) of Korea funded by the
Korean government (Ministry of Education, Science and Technology,
MEST), Grant Numbers: 2023R1A2C1008137 (Gh.-S. Y.),
RS-2025-00513982(H.-Ch.K.), and  2020R1F1A1067876 
(U. Y.). 

\bibliography{Neutron_Star}
\bibliographystyle{apsrev4-2}

\end{document}